\documentclass[showpacs,reprint,amsmath,amssymb,aps,pra]{revtex4-1}
\usepackage{graphicx}
\usepackage{dcolumn}
\usepackage{bm}
\usepackage{epstopdf}
\usepackage[mathlines]{lineno}
\usepackage{color}
\usepackage[colorlinks=true,linkcolor=blue,citecolor=blue]{hyperref}

\begin{document}
\preprint{APS/123-QED}
\title{Anomalous external-magnetic-field dependence of dephasing in a spin bath}
\author{Rui Li}
\email{rl.rueili@gmail.com}
\affiliation{Department of Physics and State Key Laboratory of
Surface Physics, Fudan University, Shanghai 200433, China}

\begin{abstract}
We theoretically investigate the dephasing of a central spin-1 model. An interesting mechanism of spin decoherence is found with this model, namely {\em hyperfine mediated spectral diffusion}. This mechanism contains both the features of dipolar interactions induced spectral diffusion and hyperfine mediated interactions. We also find an anomalous magnetic field dependence of decoherence, which is caused by the competition between crystal field splitting and Zeeman splitting of the central spin. As the external magnetic field increases, in the V type level structure regime, the decoherence rate becomes much stronger; while in the cascade type level structure regime, just like localized electron spin in quantum dots, the decoherence rate becomes much weaker.
\end{abstract}

\date{\today}
\maketitle

\section{Introduction}

The decoherence of quantum systems, due to their interaction with the surrounding environment, is a main obstacle in realizing quantum computation and quantum information processing~\cite{Nielsen, Ladd}. In order to eliminate and control the decoherence, the primary task is to understand the mechanism of decoherence in quantum systems. Electron spin is one of the promising candidates for realizing qubits in quantum computing~\cite{Loss, Hanson, Petta, Bluhm}, understanding the mechanism of spin decoherence is of great practical importance in spin based quantum computer architectures~\cite{Witzel2, RBLiu}. The central spin model $H=\gamma_{e}\,BS_{z}+\sum_{l}A_{l}\textbf{S}\cdot\textbf{I}_{l}+\sum_{l}\gamma_{n(l)}BI^{z}_{l}$ ($S = 1/2$), where a central electron spin $\textbf{S}$ interacts with a bath of nuclear spins $\textbf{I}_{l}$ via hyperfine interaction, has been studied over decades since the first investigation by Gaudin~\cite{Gaudin}. Although the model seems to be very simple, actually its dynamics is very complicated. Various methods have been developed to solve this model, including direct solution under special initial states~\cite{Khaetskii1, Khaetskii2}, the non-Markovian master equation approach~\cite{Coish1, Coish3, Barnes1, Barnes2}, the Schrieffer-Wolff-transformation-based pure dephasing model~\cite{Coish2, Cywinski1, Cywinski2, Cywinski3, Cywinski4}, numerical simulation~\cite{Dobrovitski, WXZhang1, WXZhang2}, equation of motion method~\cite{Deng1, Deng2}, etc.

At low magnetic field, the dynamics of the central spin is very complicated, both longitudinal relaxation and transverse dephasing are associated with the central spin, which makes the problem theoretically unaccessible. The numerical simulation may solve the dynamics exactly, but the underlying physics is usually unclear, and it also can only deal with a small number of bath spins~\cite{Dobrovitski, WXZhang1, WXZhang2}. At large magnetic field, the direct relaxation is forbidden due to the large energy mismatch between the central spin and the nuclear spins, so that the central spin only contains pure dephasing. With the help of Schrieffer-wolff transformation, the hyperfine mediated interactions between nuclear spins, which are responsible for the dephasing of the central spin, can be obtained~\cite{WYao, Coish2, Cywinski1, Cywinski2}.

\begin{figure}
\includegraphics[width=8.5cm]{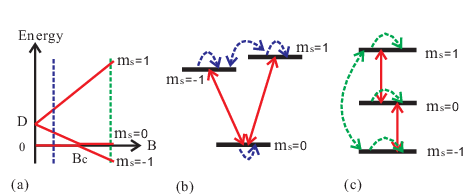}
\caption{\label{Fig_EL}(Color online) (a) The energy levels of the central electron spin as a function of external magnetic field $B$. (b) The V type level structure at weak external magnetic field $B<B_{c}$. (c) The cascade level structure at strong external magnetic field $B>B_{c}$. The solid arrows indicate the first-order relaxations, which are almost impossible due to the large energy mismatch. The dashed arrows indicate the second-order virtual relaxations, which are induced by the first-order relaxations. The relaxations from self to self imply they are pure dephasing processes.}
\end{figure}

How about when the central spin is replaced by the spin-$1$ system $H_{s}=DS^{2}_{z}+\gamma_{e}BS_{z}$ ($S = 1$)? The energy spectrum of this spin $1$ system is not as simple as that of a spin-$1/2$ system [see Fig.~\ref{Fig_EL}(a)]; there is a critical magnetic field $B_{c}=D/\gamma_{e}$, at which the states $|-\rangle$ and $|0\rangle$ are degenerate. This spin $1$ system has two types of level structure depending on the external magnetic field. At weak external magnetic field $B<B_{c}$, the level structure is V type [see Fig.~\ref{Fig_EL}(b)]; while at strong external magnetic field $B>B_{c}$, the level structure is cascade type [see Fig.~\ref{Fig_EL}(c)]. In the large level spacing area, the high-order relaxations are much richer than those in the spin-$1/2$ case, and something new is expected to appear.

The model we are investigating is related to the nitrogen-vacancy (NV) center in diamond; a central electron spin ($S=1$) interacts with the surrounding sparse ${}^{13}$C nuclear spins~\cite{Childress, Dutt, Hanson2}. However, due to the fact that the hyperfine interactions are dipolar for ${}^{13}$C nuclear spins far away from the deep defect~\cite{Gali, Maze, NZhao1, NZhao2}. In diamond with ${}^{13}$C natural abundance, the central spin only interacts with a few ${}^{13}$C nuclear spins via Fermi contact hyperfine interactions. Therefore, our model can only be applied to those diamond samples, in which at least several tens of ${}^{13}$C nuclear spins interact with the central spin via Fermi contact interactions, like in some quantum register model based on NV center spin and nearby ${}^{13}$C nuclear spins~\cite{Cappellaro} or in some ${}^{13}$C isotope enriched samples~\cite{Mizuochi}.

The paper is organized as follows. In Sec.~\ref{sec_I}, we introduce the model we are considering, and derive an effective pure dephasing Hamiltonian based on the Schrieffer-Wolff transformation. In Sec.~\ref{sec_II}, we give the detailed expression for the decoherence function based on the \emph{ring diagram} expansion technique. In Sec.~\ref{sec_III}, we give the results of dephasing under various external magnetic fields. A brief summary is given in Sec.~\ref{sec_IV}. Appendix~\ref{appendix_A} contains the detailed derivation for the effective pure dephasing Hamiltonian, and Appendix~\ref{appendix_B} contains the transformation from the Schr\"{o}dinger picture to the interaction picture for the effective time-dependent Hamiltonian.

\section{\label{sec_I}The model}

We consider a model of the NV center in diamond, where a central electron spin interacts with a bath of ${}^{13}$C nuclear spins~\cite{Childress, Dutt, Hanson2}. The NV spin is quantized along the N-V axis, an external magnetic field $B$ is also applied along this quantized direction, and the total Hamiltonian reads
\begin{equation}
H=DS^{2}_{z}+\gamma_{e}BS_{z}+\sum_{l}\textbf{S}\cdot\textbf{A}_{l}\cdot\textbf{I}_{l}+\gamma_{n}B\sum_{l}I^{z}_{l},\label{Eq_model}
\end{equation}
where $S=1$, $I=1/2$, $D=2.87$ GHz is the crystal field splitting, $\gamma_{e(n)}$ is the electron (nuclear) gyromagnetic ratio, and $\textbf{A}_{l}$ is the hyperfine coupling tensor between the $l$th nuclear spin and the electron spin. The hyperfine tensor takes the form~\cite{Gali}
\begin{eqnarray}
A^{\alpha\beta}_{l}&=&\frac{8\pi}{3}\gamma_{e}\gamma_{n}\rho_{s}(\textbf{R}_{l})\delta_{\alpha\beta}+\nonumber\\
&&\int\,d^{3}\textbf{r}\rho_{s}(\textbf{r})\gamma_{e}\gamma_{n}\frac{3r_{\alpha}r_{\beta}-\delta_{\alpha\beta}r^{2}}{r^{5}},
\end{eqnarray}
where $\rho_{s}(\textbf{r})$ is the spin density, $\textbf{R}_{l}$ is the site of the ${}^{13}$C nucleus, and $\textbf{r}$ is the displacement measured from $\textbf{R}_{l}$. The first part is the Fermi contact term, and the second part is the anisotropic dipolar term. We consider a special NV center, where the Fermi contact term contributes significantly to the hyperfine tensor for a sufficient number of ${}^{13}$C nuclear spins (in our calculation this number is chosen to be 100), so that the anisotropic dipolar term is less important and is ignored. This special NV center exists in some quantum register based samples~\cite{Cappellaro} or in some ${}^{13}$C isotope enriched samples~\cite{Mizuochi}. Therefore, we only retain the Fermi contact term for the hyperfine tensor $A^{\alpha\beta}_{l}\approx\,A_{l}\delta_{\alpha\beta}$, where $A_{l}=(8\pi/3)\gamma_{e}\gamma_{n}\rho_{s}(\textbf{R}_{l})$. Under this approximation, the model Eq.~(\ref{Eq_model}) can be reduced to the following central spin model:
\begin{eqnarray}
H&=&DS^{2}_{z}+\gamma\,BS_{z}+\sum_{l}A_{l}S_{z}I^{z}_{l}\nonumber\\
&&+\sum_{l}(A_{l}/2)(S_{+}I^{-}_{l}+S_{-}I^{+}_{l}),\label{model}
\end{eqnarray}
where $\gamma\,B$ is the effective Zeeman splitting of the electron spin, which contains the nuclei Zeeman splitting $\gamma=\gamma_{e}-\gamma_{n}$ due to the total spin $S_{z}+\sum^{N}_{l=1}I^{z}_{l}$ conserves.

We focus on the pure dephasing regime where the direct relaxations between different levels of the central spin are almost impossible due to the large energy mismatch, e.g., the two dashed lines shown in Fig.~\ref{Fig_EL}(a). The high-order hyperfine mediated interactions between nuclei are possible. With the help of the Schrieffer-Wolff transformation~\cite{Schrieffer}, we obtain an effective pure dephasing Hamiltonian (for details see Appendix~\ref{appendix_A}):
\begin{eqnarray}
H_{\eta}&\approx&\sum_{s=+,0,-}|s\rangle\langle\,s|\otimes\,H^{s}_{\eta}\nonumber\\
&=&\sum_{s=+,0,-}|s\rangle\langle\,s|\otimes\left\{\sum_{l}\Omega^{s}_{l}I^{z}_{l}+\sum_{l'\neq\,l}C^{s}_{ll'}I^{+}_{l}I^{-}_{l'}\right\},\label{PDHamiltonian}\nonumber\\
\end{eqnarray}
where $\Omega^{\pm}_{l}\approx\pm\,A_{l}$, $\Omega^{0}_{l}=\frac{\gamma\,B}{D^{2}-\gamma^{2}B^{2}}A^{2}_{l}$, $C^{\pm}_{ll'}=\frac{1}{2(D\pm\gamma\,B)}A_{l}A_{l'}$, and $C^{0}_{ll'}=-\frac{D}{D^{2}-\gamma^{2}B^{2}}A_{l}A_{l'}$.
Define $\mathcal{A}=\sum^{N}_{l=1}A_{l}$ as the total hyperfine field, our theory is valid when the following conditions
\begin{eqnarray}
2\gamma\,B&\gg&\mathcal{A},\nonumber\\
|(D^{2}-\gamma^{2}B^{2})/(\gamma\,B)|&\gg&\mathcal{A}\nonumber
\end{eqnarray}
are simultaneously satisfied. The purpose of the Schrieffer-Wolff transformation is to eliminate the first-order relaxations, which are forbidden due to the large energy mismatch between the central spin and the nuclear spins, and to elicit the second-order virtual relaxations [see Figs.~\ref{Fig_EL}(b) and \ref{Fig_EL}(c)]. The relaxations from self to self are actually pure dephasing processes. The second-order relaxation from $|+\rangle$ to $|-\rangle$ is very weak due to the level spacing between them and therefore is neglected in the current second-order approximation. In other words, only pure dephasing terms are retained in our consideration.

From the expression of $C^{s}_{ll'}$, we find there is a competition between the crystal field splitting $D$ and the Zeeman splitting $\gamma\,B$, which induces the two level structures of the central electron spin (see Fig.~\ref{Fig_EL}). In the weak external magnetic field $D\gg\,\gamma\,B$ regime, i.e., the V type level structure regime [see Fig.~\ref{Fig_EL}(b)], $C^{+}_{ll'}\approx\,C^{-}_{ll'}\approx\frac{A_{l}A_{l'}}{2D}$, projecting the Hamiltonian into the space spanned by $|+\rangle$ and $|-\rangle$, and we obtain
\begin{equation}
H_{\eta}=\sum_{l}A_{l}\sigma_{z}I^{z}_{l}+\sum_{l'\neq\,l}\frac{A_{l}A_{l'}}{2D}I^{+}_{l}I^{-}_{l'},\label{FID}
\end{equation}
where $\sigma_{z}=|+\rangle\langle+|-|-\rangle\langle-|$. Something unusual happens, unlike in the spin-1/2 case, the hyperfine mediated interactions are not $\sigma_{z}$ conditioned~\cite{Coish2, Cywinski1}. The hyperfine mediated interactions can introduce fluctuations to the Overhauser field; we name this dephasing as {\em hyperfine mediated spectral diffusion}. This mechanism of dephasing is different from the two main mechanisms in quantum dots, i.e., dipolar spectral diffusion~\cite{Sousa1, Sousa2, Witzel2, Witzel1} and hyperfine mediated interactions~\cite{WYao, Coish2, Cywinski1}. The {\em hyperfine mediated spectral diffusion} somehow comprises both the features of these two mechanisms; e.g., the hyperfine mediated interactions are long-range and introduce fluctuations to the Overhauser field.

In the strong external magnetic field $D\ll\,\gamma\,B$ regime, i.e., the level structure of the central electron spin is cascade type [see Fig.~\ref{Fig_EL}(c)]. The coefficients  $C^{\pm}_{ll'}\approx\pm\frac{A_{l}A_{l'}}{2\gamma\,B}$, projecting the Hamiltonian into the space spanned by $|+\rangle$ and $|-\rangle$, and we obtain
\begin{equation}
H_{\eta}=\sum_{l}A_{l}\sigma_{z}I^{z}_{l}+\sum_{l'\neq\,l}\frac{A_{l}A_{l'}}{2\gamma\,B}\sigma_{z}I^{+}_{l}I^{-}_{l'}.
\end{equation}
Just like in the-spin 1/2 case, the hyperfine mediated interactions are $\sigma_{z}$ conditioned~\cite{Coish2, Cywinski1}. There is no decoherence under spin echo with this case~\cite{WYao, Cywinski1}. The discussion is similar in the space spanned by $|+\rangle$ and $|0\rangle$ or $|-\rangle$ and $|0\rangle$.

\section{\label{sec_II}Decoherence function}

For free induction decay, the dominant dephasing source is the Overhauser field, i.e., the inhomogeneous broadening to the spin. The decoherence function is defined as the normalized off-diagonal element of the reduced density matrix of the central spin~\cite{WYao, Cywinski2}
\begin{equation}
\rho_{+,-}(t)=\langle\,e^{iH^{-}_{\eta}t}e^{-iH^{+}_{\eta}t}\rangle\approx\langle\,e^{i\sum_{l}\Omega^{-}_{l}I^{z}_{l}t}e^{-i\sum_{l}\Omega^{+}_{l}I^{z}_{l}t}\rangle,
\end{equation}
where $\langle\cdots\rangle=\mathrm{Tr}_{B}\{\rho_{B}(0)\cdots\}$ denotes the average over the configuration of bath spins, with $\rho_{B}(0)=(1/2^{N})\textbf{1}_{1}\otimes\cdots\textbf{1}_{N}$ being the initial unpolarized bath density matrix.
One can easily obtain $\rho_{+,-}(t)=\prod_{l}\cos(A_{l}t)$ and $\rho_{+,0}(t)=\prod_{l}\cos(A_{l}t/2)$. This Overhauser field induced decay is shown in Fig.~\ref{FID}; the decay takes a Gauss shape and $\rho_{+,-}(t)$ decays faster than $\rho_{+,0}(t)$~\cite{NZhao1}. This dephasing can be largely suppressed by a Hahn spin echo~\cite{Hahn}, which leads us to investigate spin echo decay. Under spin echo, the decoherence function takes the form~\cite{Witzel2, Cywinski2}
\begin{equation}
W_{+,-}(t)=\langle\,e^{iH^{+}_{\eta}t/2}e^{iH^{-}_{\eta}t/2}e^{-iH^{+}_{\eta}t/2}e^{-iH^{-}_{\eta}t/2}\rangle.
\end{equation}
It should be noted that, there are three different off-diagonal elements for the central spin reduced density matrix; correspondingly, there are three different decoherence functions, the other two are $W_{+,0}(t)$ and $W_{-,0}(t)$.

\begin{figure}
\includegraphics[width=8.5cm]{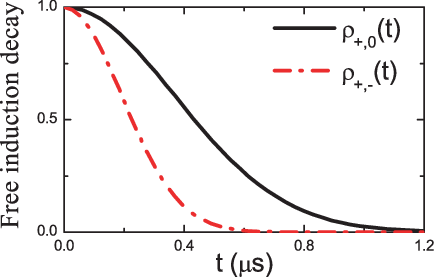}
\caption{\label{FID}(Color online) Free induction decay. The decoherence is dominated by the Overhauser field. The hyperfine coupling is randomly chosen as $A_{l}=0.5+0.1\times\mathrm{randn}(1,N)$ MHz, where $N=100$ is the total number of nuclear spins.}
\end{figure}
\begin{figure}
\includegraphics{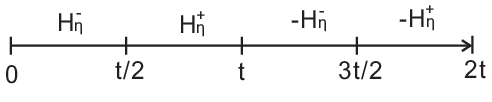}
\caption{\label{Fig_H}The Hamiltonians in different domains of evolution time.}
\end{figure}

The sequence of evolution operators in the decoherence function $W(t)$ can be considered as an effective time dependent Hamiltonian evolving from $0\rightarrow\,2t$ (see Fig.~\ref{Fig_H})
\begin{equation}
H_{ef}(\tau)=-\alpha(2t;\tau)H^{-}_{\eta}+\beta(2t;\tau)H^{+}_{\eta},\label{effective_H}
\end{equation}
where $\alpha(2t;\tau)=-\theta(\frac{t}{2}-\tau)\theta(\tau)+\theta(\frac{3}{2}t-\tau)\theta(\tau-t)$, and
$\beta(2t;\tau)=\theta(t-\tau)\theta(\tau-\frac{t}{2})-\theta(2t-\tau)\theta(\tau-\frac{3}{2}t)$,
with $\theta(\tau)$ being the usual step function. The decoherence function is correspondingly written as
\begin{equation}
W_{+,-}(t)=\langle\mathcal{T}e^{-i\int^{2t}_{0}d\tau\,H_{ef}(\tau)}\rangle.
\end{equation}
The effective time dependent Hamiltonian $H_{ef}(\tau)$ chosen here is a little different from the previous treatment~\cite{Cywinski2}, in which an effective Hamiltonian is only associated with the evolution from $0\rightarrow\,t$. The advantages of our treatment are that the elusory Keldysh contour will not appear any more and the subsequent calculations are much simpler. We transform into the interaction picture with respect to the exchange interactions in $H_{ef}$ (see Appenix~\ref{appendix_B})
\begin{equation}
H_{int}(\tau)=\sum_{l'\neq\,l}f_{ll'}(2t;\tau)I^{+}_{l}I^{-}_{l'},
\end{equation}
where
\begin{eqnarray}
f_{ll'}(2t;\tau)&=&\big[C^{+}_{ll'}\beta(2t;\tau)-C^{-}_{ll'}\alpha(2t;\tau)\big]\times\nonumber\\
&&e^{i\big[(\Omega^{+}_{l}-\Omega^{+}_{l'})\tilde{\beta}(2t;\tau)-(\Omega^{-}_{l}-\Omega^{-}_{l'})\tilde{\alpha}(2t;\tau)\big]},
\end{eqnarray}
with $\tilde{\alpha}(2t;\tau)=\int^{\tau}_{0}d\tau'\alpha(2t;\tau')$ and $\tilde{\beta}(2t;\tau)=\int^{\tau}_{0}d\tau'\beta(2t;\tau')$. The evolution operators between the Schr\"{o}dinger picture and the interaction picture are related by $U_{ef}(2t)=U_{0}(2t)U_{int}(2t)$. For spin echo, $U_{0}(2t)=1$~\cite{Cywinski2}, so that the decoherence function has a compact form:
\begin{equation}
W_{+,-}(t)=\langle\mathcal{T}e^{-i\int^{2t}_{0}d\tau\,H_{int}(\tau)}\rangle.
\end{equation}
The above expression is very similar to the thermodynamic potential in quantum field theory~\cite{Mahan}; however, there are also some differences due to the properties of the spin operator, e.g., the commutation relation of the spin operator is $[I^{+}_{l},I^{-}_{l'}]=2\delta_{ll'}I^{z}_{l}$, and Wick's theorem is not applicable. The linked-cluster expansion approach~\cite{Saikin} and the cluster-correlation expansion approach~\cite{WYang1, WYang2} were developed to deal with this kind of problem, and great progress was made by Cywinski \emph{et al}. by using the \emph{ring diagram} expansion technique~\cite{Cywinski2}. The \emph{ring diagrams} $R_{k}(t)$ are defined as~\cite{Cywinski2}
\begin{equation}
R_{k}(t)=\sum_{\substack{l_{1}\neq\,l_{2},l_{2}\neq\,l_{3}\\\cdots,l_{k}\neq\,l_{1}}}T_{l_{1}l_{2}}(t)T_{l_{2}l_{3}}(t)\cdots\,T_{l_{k}l_{1}}(t),
\end{equation}
where the $T$ matrix
\begin{eqnarray}
T_{ll'}&=&\frac{1-\delta_{ll'}}{2}\int^{2t}_{0}d\tau\,f_{ll'}(2t;\tau)\nonumber\\
&=&(1-\delta_{ll'})\times2i\left(\frac{C^{+}_{ll'}}{\Omega^{+}_{ll'}}-\frac{C^{-}_{ll'}}{\Omega^{-}_{ll'}}\right)\times\nonumber\\
&&\sin(\frac{\Omega^{+}_{ll'}t}{4})\sin(\frac{\Omega^{-}_{ll'}t}{4})\mathrm{exp}\left(i\frac{(\Omega^{+}_{ll'}+\Omega^{-}_{ll'})t}{4}\right),
\end{eqnarray}
with $\Omega^{\pm}_{ll'}=\Omega^{\pm}_{l}-\Omega^{\pm}_{l'}$. The decoherence function can be obtained with \emph{ring diagram} expansion~\cite{Cywinski2}
\begin{equation}
W_{+,-}(t)=\mathrm{exp}\left(\sum^{\infty}_{k=2}\frac{(-i)^{k}}{k}R_{k}(t)\right).
\end{equation}
It should be noted that all the formulas derived so far are very general. They are universal to the phase decoherence between two arbitrary states; e.g., for the phase decoherence $W_{+,0}(t)$ between $|+\rangle$ and $|0\rangle$, we just replace the symbol '$-$' with '$0$' in all the above formulas.

\section{\label{sec_III}Magnetic field dependence of echo decay}

From the expressions of $C^{s}_{ll'}$ in the Hamiltonian [see Eq.~(\ref{PDHamiltonian})], we know that the decoherence has a magnetic field dependence. In our calculation, the total number of nuclear spins is chosen as $N=100$, and the hyperfine couplings $A_{l}$ are randomly and canonically distributed around an average value $\langle\,A_{l}\rangle=\sum^{N}_{l=1}A_{l}/N=0.5$ MHz. Actually, the physics under study has nothing to do with these specific choices of hyperfine couplings. The magnetic field dependence is determined by the ratio $\alpha=\gamma\,B/D$.

\begin{figure}
\includegraphics[width=8.5cm]{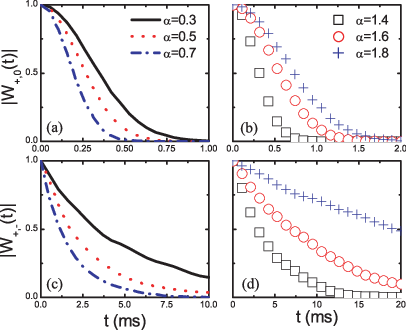}
\caption{\label{Fig_magnetic}(Color online) The magnetic-field dependence of echo decay. The hyperfine couplings are chosen as $A_{l}=0.5+0.1\times\,\mathrm{randn}(1,N)$ MHz, where $N$ is the total number of nuclear spins; we choose $N=100$ here. We use $\alpha=\gamma\,B/D$ to mark the magnetic-field dependence. [(a), (c)] Echo decay of central spin in V type level structure regime. [(b), (d)] Echo decay of central spin in the cascade level structure regime.}
\end{figure}

In quantum dots, when the applied magnetic field increases, the decoherence rate is linearly decreasing~\cite{Cywinski2}. The reason is simple: the larger the Zeeman splitting is, the weaker the hyperfine mediated interactions become, and so the decoherence rate becomes weaker. The case is a little different here for the NV spin (see Fig.~\ref{Fig_magnetic}). As the external magnetic field increases, in the V type level structure regime, the decoherence rate becomes much stronger [see Figs.~\ref{Fig_magnetic}(a) and \ref{Fig_magnetic}(c)]; while in the cascade level structure regime, the decoherence rate becomes much weaker [see Figs.~\ref{Fig_magnetic}(b) and \ref{Fig_magnetic}(d)]. The magnetic field dependence of decoherence is anomalous and counterintuitive in the V type level structure regime. As the external magnetic field increases, the level spacing between $|+\rangle$ and $|0\rangle$ also increases. One may imagine the hyperfine mediated interactions will also become weaker, and thus the decoherence rate should be linearly decreasing. This physical picture is true only in the absence of the third level $|-\rangle$. The change of the third level $|-\rangle$ with the magnetic field influences the decoherence between $|+\rangle$ and $|0\rangle$. Especially in the vicinity of critical magnetic field $B_{c}=D/\gamma$ $(\alpha=1)$, the relaxation from $|0\rangle$ to $|-\rangle$ is very strong and the influence of the third level $|-\rangle$ becomes much more evident. In the cascade level structure regime, the magnetic-field dependence of decoherence is normal, just like the electron spin-$1/2$ in quantum dots, and the decoherence rate is inversely proportional to the external magnetic field.

\begin{figure}
\includegraphics[width=8.5cm]{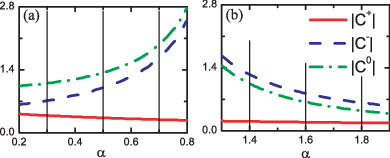}
\caption{\label{Fig_coefficients}(Color online) The hyperfine mediated coupling coefficients as a function of the external magnetic field. (a) V type level structure regime. (b) Cascade level structure regime.}
\end{figure}

All of these effects can be well understood by investigating the magnetic field dependence of the hyperfine mediated interactions, where  $C^{\pm}_{ll'}=\frac{A_{l}A_{l'}}{D}C^{\pm}$ and $C^{0}_{ll'}=-\frac{A_{l}A_{l'}}{D}C^{0}$, with the coefficients $C^{\pm}=1/2(1\pm\alpha)$ and $C^{0}=1/(1-\alpha^{2})$. The echo signals are proportional to the magnitude of these coefficients. The variations of these coefficients with magnetic field reflect the variations of echo decay. We plot the magnetic-field dependence of $|C^{\pm}|$ and $|C^{0}|$ in the validity regime of our theory (see Fig.~\ref{Fig_coefficients}). The variations of $|C^{-}|$ and $|C^{0}|$ are opposite between the two level structure regimes, and they are more sensitive to the magnetic field than $|C^{+}|$, thus the anomalous magnetic field dependence of decoherence appears. It should be noted that this explanation has nothing to do with the detailed hyperfine couplings $A_{l}$, it is completely due to the competition between $D$ and $\gamma\,B$, which induces the two different level structures.

\section{\label{sec_IV}Summary}

We have theoretically investigated the hyperfine mediated interaction induced dephasing of a central spin ($S=1$) model. In a large level spacing area, the effective pure dephasing Hamiltonian is obtained with the help of the Schrieffer-Wolff transformation. An interesting mechanism of spin dephasing is found with this model, namely \emph{hyperfine mediated spectral diffusion}. In this mechanism, the fluctuations of the Overhauser field come from hyperfine mediated interactions. This dephasing mechanism is different from the dipolar spectral diffusion~\cite{Sousa1, Sousa2, Witzel2, Witzel1} and the hyperfine mediated interactions~\cite{WYao, Coish2, Cywinski1}; \emph{hyperfine mediated spectral diffusion} contains both the features of these two mechanisms.  We also investigated the external magnetic dependence of echo decay. The magnetic field dependence of decoherence is anomalous in the V type level structure regime; the decoherence rate is proportional to the external magnetic field. While in the cascade level structure regime, just like electron spin-$1/2$ in quantum dots, the decoherence rate is inversely proportional to the external magnetic field. The physical effects we studied here may exist in some NV center samples.

\acknowledgements
We thank J.Q. You, Wenxian Zhang, and Xuedong Hu for discussions. This work is supported by the National Basic Research Program of China under Grant No. 2009CB929300 and the National Natural Science Foundation of China under Grant No. 91121015.

\appendix
\section{\label{appendix_A}Derivation of the effective Hamiltonian}
In this appendix, we give the detailed derivation for the effective Hamiltonian. The method used here is based on the Schrieffer-Wolff transformation~\cite{Schrieffer}; this method also has been used to derive the hyperfine mediated interactions between nuclei in quantum dots~\cite{Coish2, Cywinski1}. Due to the energy landscape for the spin-$1$ system, the transformation here is a little more complicated than in quantum dots. We give the details for interested readers. We write the Hamiltonian again
\begin{eqnarray}
H&=&H_{0}+H_{1},\nonumber\\
H_{0}&=&DS^{2}_{z}+\gamma\,BS_{z}+\sum_{l}A_{l}S_{z}I^{z}_{l},\nonumber\\
H_{1}&=&\sum_{l}(A_{l}/2)(S_{+}I^{-}_{l}+S_{-}I^{+}_{l}).
\end{eqnarray}
We only focus on the large level spacing area, where the direct relaxations between different energy levels are almost impossible due to the large energy mismatch between the central spin and the nuclear spins. The Schrieffer-Wolff transformation states that~\cite{Schrieffer} one can make a canonical transformation to the Hamiltonian
\begin{eqnarray}
H_{\eta}=e^{\eta}He^{-\eta}&=&H_{0}+[\eta,H_{0}]+H_{1}+\frac{1}{2}\big[\eta,[\eta,H_{0}]+H_{1}\big]\nonumber\\
&&+\frac{1}{2}[\eta,H_{1}]+\mathcal{O}(H^{3}_{1}).
\end{eqnarray}
When $\eta$ satisfy
\begin{equation}
[\eta,H_{0}]+H_{1}=0,\label{S_eta}
\end{equation}
up to only the second-order terms of $H_{1}$, one obtains the effective Hamiltonian
\begin{equation}
H_{\eta}=H_{0}+\frac{1}{2}[\eta,H_{1}].\label{eff_H}
\end{equation}
Thus, the essential problem of deriving the effective Hamiltonian is to solve $\eta$. For our model, we give an ansatz
\begin{equation}
\eta=\sum_{l}\left((a^{1}_{l}\tilde{S}_{+}+a^{2}_{l}S_{+})I^{-}_{l}-(b^{1}_{l}\tilde{S}_{-}+b^{2}_{l}S_{-})I^{+}_{l}\right),\label{eta}
\end{equation}
where $a^{1}_{l}$, $a^{2}_{l}$, $b^{1}_{l}$, and $b^{2}_{l}$ are the coefficients to be determined, and
\begin{eqnarray}
\tilde{S}_{+}&=&S_{z}S_{+}+S_{+}S_{z},\nonumber\\
\tilde{S}_{-}&=&S_{-}S_{z}+S_{z}S_{-}.
\end{eqnarray}
When the $\eta$ in Eq.~(\ref{S_eta}) is replaced by Eq.~(\ref{eta}), we obtain the equation array for the coefficients
\begin{eqnarray}
&&(D-A_{l}/2)a^{1}_{l}+\big(\gamma(B+B_{o})+A_{l}/2\big)a^{2}_{l}=A_{l}/2,\nonumber\\
&&(D-A_{l}/2)b^{1}_{l}+\big(\gamma(B+B_{o})-A_{l}/2\big)b^{2}_{l}=A_{l}/2,\nonumber\\
&&(D-A_{l}/2)a^{2}_{l}+\big(\gamma(B+B_{o})+A_{l}/2\big)a^{1}_{l}=0,\nonumber\\
&&(D-A_{l}/2)b^{2}_{l}+\big(\gamma(B+B_{o})-A_{l}/2\big)b^{1}_{l}=0,\nonumber\\
\end{eqnarray}
where $\gamma\,B_{o}=\sum^{N}_{l'=1}A_{l'}I^{z}_{l'}$ is the Overhauser field. The coefficients can be simply solved by
\begin{eqnarray}
a^{1}_{l}&=&\frac{D-A_{l}/2}{(D-A_{l}/2)^{2}-\big(\gamma(B+B_{o})+A_{l}/2\big)^{2}}\times\frac{A_{l}}{2},\nonumber\\
a^{2}_{l}&=&\frac{-\big(\gamma(B+B_{o})+A_{l}/2\big)}{(D-A_{l}/2)^{2}-\big(\gamma(B+B_{o})+A_{l}/2\big)^{2}}\times\frac{A_{l}}{2},\nonumber\\
b^{1}_{l}&=&\frac{D-A_{l}/2}{(D-A_{l}/2)^{2}-\big(\gamma(B+B_{o})-A_{l}/2\big)^{2}}\times\frac{A_{l}}{2},\nonumber\\
b^{2}_{l}&=&\frac{-\big(\gamma(B+B_{o})-A_{l}/2\big)}{(D-A_{l}/2)^{2}-\big(\gamma(B+B_{o})-A_{l}/2\big)^{2}}\times\frac{A_{l}}{2}.\nonumber\\
\end{eqnarray}
Define $\mathcal{A}=\sum^{N}_{l=1}A_{l}$ as the total hyperfine field, then $\mathcal{A}\geq2\gamma\,B_{o}$. We focus on the magnetic-field regime where both the following conditions
\begin{eqnarray}
2\gamma\,B&\gg&\mathcal{A},\nonumber\\
|(D^{2}-\gamma^{2}B^{2})/(\gamma\,B)|&\gg&\mathcal{A}\nonumber
\end{eqnarray}
are simultaneously satisfied, and then the coefficients can be simplified as
\begin{eqnarray}
a^{1}_{l}&\approx&b^{1}_{l}\approx\,DA_{l}/2(D^{2}-\gamma^{2}B^{2}),\nonumber\\
a^{2}_{l}&\approx&b^{2}_{l}\approx-\gamma\,BA_{l}/2(D^{2}-\gamma^{2}B^{2}).
\end{eqnarray}
Once $\eta$ is obtained, the effective Hamiltonian can be evaluated by Eq.~(\ref{eff_H}):
\begin{eqnarray}
H_{\eta}&=&DS^{2}_{z}+\gamma\,BS_{z}+\sum_{l}A_{l}(1-a^{1}_{l})S_{z}I^{z}_{l}\nonumber\\
&&+\sum_{l'\neq\,l}A_{l}a^{1}_{l'}\left(
                                  \begin{array}{ccc}
                                    1 & 0 & 0 \\
                                    0 & -2 & 0 \\
                                    0 & 0 & 1 \\
                                  \end{array}
                                \right)I^{+}_{l}I^{-}_{l'}\nonumber\\
&&-\sum_{l}A_{l}a^{2}_{l}\left(
\begin{array}{ccc}
1&0&0\\
0&2&0\\
0&0&1\\
 \end{array}
 \right)I^{z}_{l}+\sum_{l'\neq\,l}A_{l}a^{2}_{l'}S_{z}I^{+}_{l}I^{-}_{l'}\nonumber\\
 &&+\sum_{l'\neq\,l}\frac{A_{l}a^{1}_{l'}}{2}\{S^{2}_{+}I^{-}_{l}I^{-}_{l'}+S^{2}_{-}I^{+}_{l}I^{+}_{l'}\}.
\end{eqnarray}
The last term in above equation is the virtual second-order relaxation from $|1\rangle\rightarrow|-1\rangle$. Due to the level splitting between $|1\rangle$ and $|-1\rangle$, the relaxation is very weak, therefore, we can neglect it in our consideration. Thus, we obtain an effective pure dephasing Hamiltonian
\begin{eqnarray}
H_{\eta}&\approx&\sum_{s=+,0,-}|s\rangle\langle\,s|\otimes\left\{\sum_{l}\Omega^{s}_{l}I^{z}_{l}+\sum_{l'\neq\,l}C^{s}_{ll'}I^{+}_{l}I^{-}_{l'}\right\},\nonumber\\
\end{eqnarray}
where $\Omega^{\pm}_{l}=-A_{l}a^{2}_{l}\pm\,A_{l}(1-a^{1}_{l})\approx\pm\,A_{l}$, $\Omega^{0}_{l}=-2A_{l}a^{2}_{l}=\frac{\gamma\,B}{D^{2}-\gamma^{2}B^{2}}A^{2}_{l}$, $C^{\pm}_{ll'}=A_{l}(a^{1}_{l'}\pm\,a^{2}_{l'})=\frac{1}{2(D\pm\gamma\,B)}A_{l}A_{l'}$, and $C^{0}_{ll'}=-2A_{l}a^{1}_{l'}=-\frac{D}{D^{2}-\gamma^{2}B^{2}}A_{l}A_{l'}$.

\section{\label{appendix_B}Interaction picture}
In this appendix, we give the detailed derivation of the Hamiltonian in the interaction picture. We write $H_{ef}$ as
\begin{eqnarray}
H_{ef}(\tau)&=&H_{0}(\tau)+H_{1}(\tau),\nonumber\\
H_{0}(\tau)&=&\sum_{l}\big[\Omega^{+}_{l}\beta(2t;\tau)-\Omega^{-}_{l}\alpha(2t;\tau)\big]I^{z}_{l},\nonumber\\
H_{1}(\tau)&=&\sum_{l'\neq\,l}\big[C^{+}_{ll'}\beta(2t;\tau)-C^{-}_{ll'}\alpha(2t;\tau)\big]I^{+}_{l}I^{-}_{l'}.
\end{eqnarray}
Although $H_{0}(\tau)$ is time dependent, the commutation $[H_{0}(\tau),H_{0}(\tau')]=0$ still holds, so there is no need to apply the time ordering operator $\mathcal{T}$ to the time evolution operator $U_{0}(\tau)$
\begin{eqnarray}
U_{0}(\tau)&=&\mathcal{T}\mathrm{exp}\left(-i\int^{\tau}_{0}d\tau'H_{0}(\tau')\right)\nonumber\\
&=&\mathrm{exp}\left(-i\sum_{l}\big[\Omega^{+}_{l}\tilde{\beta}(2t;\tau)-\Omega^{-}_{l}\tilde{\alpha}(2t;\tau)\big]I^{z}_{l}\right),\nonumber\\
\end{eqnarray}
where
\begin{eqnarray}
\tilde{\alpha}(2t;\tau)&=&\int^{\tau}_{0}d\tau'\alpha(2t;\tau'),\nonumber\\
\tilde{\beta}(2t;\tau)&=&\int^{\tau}_{0}d\tau'\beta(2t;\tau').
\end{eqnarray}
Transforming into the interaction picture with respect to $H_{1}(\tau)$, we obtain
\begin{equation}
H_{int}(\tau)=U^{+}_{0}(\tau)H_{1}(\tau)U_{0}(\tau)=\sum_{l'\neq\,l}f_{ll'}(2t;\tau)I^{+}_{l}I^{-}_{l'},
\end{equation}
where
\begin{eqnarray}
f_{ll'}(2t;\tau)&=&\big[C^{+}_{ll'}\beta(2t;\tau)-C^{-}_{ll'}\alpha(2t;\tau)\big]\nonumber\\
&&\times\,e^{i\big[(\Omega^{+}_{l}-\Omega^{+}_{l'})\tilde{\beta}(2t;\tau)-(\Omega^{-}_{l}-\Omega^{-}_{l'})\tilde{\alpha}(2t;\tau)\big]}.
\end{eqnarray}
We want to emphasize here that although the expression is derived in the space spanned by $|+\rangle$ and $|-\rangle$, actually it is universal for two arbitrary states; e.g., for the space spanned by $|+\rangle$ and $|0\rangle$, we just replace the symbol '$0$' for '$-$' in the above expression.

\end{document}